\documentclass[useAMS,usenatbib]{mnras}
\usepackage{showyourwork}

\usepackage{natbib}
\usepackage{tabu,booktabs}
\usepackage{amsmath}
\usepackage{xcolor}
\usepackage[english]{babel}
\usepackage{graphicx}
\usepackage{xspace}

\usepackage{longtable}
\usepackage{threeparttablex}
\usepackage{array}

\makeatletter
\def\abstract{\if@twocolumn
   \start@SFBbox\@abstract
 \else
   \@abstract
 \fi}
\def\endabstract{\if@twocolumn
   \endlist\finish@SFBbox
 \else
  \endlist
 \fi}
\makeatother
\newcommand{\hMpc}{{\ifmmode{h^{-1}{\rm Mpc}}\else{$h^{-1}$Mpc}\fi}}
\newcommand{\Mpc}{{\ifmmode{{\rm Mpc}}\else{Mpc}\fi}}
\newcommand{\hkpc}{{\ifmmode{h^{-1}{\rm kpc}}\else{$h^{-1}$kpc}\fi}}
\newcommand{\kpc}{{\ifmmode{ {\rm kpc} }\else{{\rm kpc}}\fi}}
\newcommand{\kms}{{\ifmmode{ {\rm km\,s^{-1}} }\else{ ${\rm km\,s^{-1}}$ }\fi}}
\newcommand{\hMsun}{{\ifmmode{h^{-1}{\rm {M_{\odot}}}}\else{$h^{-1}{\rm{M_{\odot}}}$}\fi}}
\newcommand{\Msun}{{\ifmmode{{\rm M}_{\odot}}\else{${\rm M}_{\odot}$}\fi}}
\newcommand{\Mhalo}{{\ifmmode{M_{\rm halo}}\else{$M_{\rm halo}$}\fi}}
\newcommand{\Rvir}{{\ifmmode{R_{\rm vir}}\else{$R_{\rm vir}$}\fi}}
\newcommand{\Mvir}{{\ifmmode{M_{\rm vir}}\else{$M_{\rm vir}$}\fi}}
\newcommand{\Mstar}{{\ifmmode{M_{\rm star}}\else{$M_{\rm star}$}\fi}}
\newcommand{\Vrot}{{\ifmmode{V_{\rm rot}}\else{$V_{\rm rot}$}\fi}}
\newcommand{\ltsima}{$\; \buildrel < \over \sim \;$}
\newcommand{\gtsima}{$\; \buildrel > \over \sim \;$}
\newcommand{\lsim}{\lower.5ex\hbox{\ltsima}}
\newcommand{\gsim}{\lower.5ex\hbox{\gtsima}}

\def\lesssim{\mathrel{\hbox{\rlap{\hbox{\lower4pt\hbox{$\sim$}}}\hbox{$<$}}}}
\def\gtrsim{\mathrel{\hbox{\rlap{\hbox{\lower4pt\hbox{$\sim$}}}\hbox{$>$}}}}

\newcommand{\beq}{\begin{equation}}
\newcommand{\eeq}{\end{equation}}
\def\beqa{\begin{eqnarray}}
\def\eeqa{\end{eqnarray}}
\def\LCDM{\ensuremath{\Lambda}CDM}

\def\head{ \vbox to 0pt{\vss \hbox to 0pt{\hskip 440pt\rm
      LA-UR-10-07069\hss} \vskip 25pt}}

\def \kms {\ifmmode  \,\rm km\,s^{-1} \else $\,\rm km\,s^{-1}  $ \fi }
\def \kpc {\ifmmode  {\rm kpc}  \else ${\rm  kpc}$ \fi  }  
\def \hkpc {\ifmmode  {h^{-1}\rm kpc}  \else ${h^{-1}\rm kpc}$ \fi  }  
\def \hMpc {\ifmmode  {h^{-1}\rm Mpc}  \else ${h^{-1}\rm Mpc}$ \fi  }  
\def \Mpch {\ifmmode  {h^{-1}\rm Mpc}  \else ${h^{-1}\rm Mpc}$ \fi  }  
\def \Msun {\ifmmode {\rm M}_{\odot} \else ${\rm M}_{\odot}$ \fi} 
\def \hMsun {\ifmmode h^{-1}\,\rm M_{\odot} \else $h^{-1}\,\rm M_{\odot}$ \fi}

\def \LCDM {\ifmmode \Lambda{\rm CDM} \else $\Lambda{\rm CDM}$ \fi}
\def \sig8 {\ifmmode \sigma_8 \else $\sigma_8$ \fi} 
\def \OmegaM {\ifmmode \Omega_{\rm m} \else $\Omega_{\rm m}$ \fi} 
\def \Omegab {\ifmmode \Omega_{\rm b} \else $\Omega_{\rm b}$ \fi} 
\def \OmegaL {\ifmmode \Omega_{\rm \Lambda} \else $\Omega_{\rm \Lambda}$\fi} 
\def \Deltavir {\ifmmode \Delta_{\rm vir} \else $\Delta_{\rm vir}$ \fi}
\def \rhocrit {\ifmmode \rho_{\rm crit} \else $\rho_{\rm crit}$ \fi}
\def \rhou {\ifmmode \rho_{\rm u} \else $\rho_{\rm u}$ \fi}
\def \zc {\ifmmode z_{\rm c} \else $z_{\rm c}$ \fi}

\def\aj{AJ}%
\def\araa{ARA\&A}%
\def\apj{ApJ}%
\def\apjl{ApJ}%
\def\apjs{ApJS}%
\def\ao{Appl.~Opt.}%
\def\aap{A\&A}%
\def\aapr{A\&A~Rev.}%
\def\mnras{MNRAS}%
\def\pasp{PASP}%
\def\nat{Nature}%
\def\head{ .ps \vbox to 0pt{\vss \hbox to 0pt{\hskip 440pt\rm
      LA-UR-10-07069\hss} \vskip 25pt}} 

\def \spose#1{\hbox  to 0pt{#1\hss}}  
\def \lta{\mathrel{\spose{\lower 3pt\hbox{$\sim$}}\raise 2.0pt\hbox{$<$}}}
\def \gta{\mathrel{\spose{\lower 3pt\hbox{$\sim$}}\raise 2.0pt\hbox{$>$}}}

\title[Early massive accretion events in MW-mass galaxies]{The impact of early massive mergers on the chemical evolution of Milky Way-like galaxies: insights from NIHAO-UHD simulations}
\author[T. Buck] {Tobias Buck$^{1,2}$\thanks{E-mail: tobias.buck@iwr-uni-heidelberg.de.de}, Aura Obreja$^{3}$, Bridget Ratcliffe$^{4}$, Yuxi(Lucy) Lu$^{5, 6}$, 
\newauthor Ivan Minchev$^{4}$, Andrea V. Macci\`o$^{7,8,9}$ \\  %, \etal \\
$^1$Universit\"at Heidelberg, Interdisziplin\"ares Zentrum f\"ur Wissenschaftliches Rechnen, Im Neuenheimer Feld 205, \\ D-69120 Heidelberg, Germany\\
$^2$Universit\"at Heidelberg, Zentrum f\"ur Astronomie, Institut f\"ur Theoretische Astrophysik, Albert-Ueberle-Straße 2, \\ D-69120 Heidelberg, Germany\\
$^3$Universit\"ats-Sternwarte M\"unchen, Scheinerstraße 1, D-81679 M\"unchen, Germany\\
${4}$Leibniz Institute for Astrophysics Potsdam, An der Sternwarte 16, 14482 Potsdam, Germany\\
$^{5}$Department of Astronomy, Columbia University, 550 West 120$^{\rm th}$ Street, New York, NY, USA\\
$^{6}$American Museum of Natural History, Central Park West, Manhattan, NY, USA\\
$^7$New York University Abu Dhabi, PO Box 129188, Saadiyat Island, Abu Dhabi, United Arab Emirates\\
$^8$Center for Astro, Particle and Planetary Physics (CAP3), New York University Abu Dhabi\\
$^9$Max-Planck-Institut f\"ur Astronomie, K\"onigstuhl 17, 69117 Heidelberg, Germany
}

\begin{document}

\date{Accepted XXXX . Received XXXX; in original form XXXX}

\pagerange{\pageref{firstpage}--\pageref{lastpage}} \pubyear{2022}

\maketitle

\label{firstpage}

\begin{abstract}
Recent observations of the Milky Way (MW) found an unexpected steepening of the star-forming gas metallicity gradient around the time of the Gaia-Sausage-Enceladus (GSE) merger event.
Here we investigate the influence of early ($t_{\rm{merger}}\lesssim5$ Gyr) massive ($M_{\rm{gas}}^{\rm{merger}}/M_{\rm{gas}}^{\rm{main}}(t_{\rm{merger}})\gsim10\%$) merger events such as the Gaia-Sausage Enceladus merger in the MW on the evolution of the cold gas metallicity gradient. We use the NIHAO-UHD suite of cosmological hydrodynamical simulations of MW-mass galaxies to study the frequency of massive early mergers and their detailed impact on the morphology and chemistry of the gaseous disks. 
We find a strong steepening of the metallicity gradient at early times for all four galaxies in our sample which is caused by a sudden increase in the cold gas disk size (up to a factor of 2) in combination with the supply of un-enriched gas ($\sim0.75$ dex lower compared to the main galaxy) by the merging dwarf galaxies.
The mergers mostly affect the galaxy outskirts and lead to an increase in cold gas surface density of up to 200\% outside of $\sim8$ kpc. The addition of un-enriched gas breaks the self-similar enrichment of the inter-stellar-medium and causes a dilution of the cold gas in the outskirts of the galaxies. 
The accreted stars and the ones formed later out of the accreted gas inhabit distinct tracks offset to lower [$\alpha$/Fe] and [Fe/H] values compared to the main galaxy's stars. We find that such mergers can contribute significantly to the formation of a second, low-$\alpha$ sequence as is observed in the MW.
\end{abstract}

%%%%%%%%%%%%%%%%%%%%%%%%%%%%%%%%%%%%%%%%%%%%%%%%%%%
\noindent
\begin{keywords}

Galaxy: structure --- Galaxy: evolution --- Galaxy: kinematics and dynamics --- galaxies:  formation --- galaxies: ISM --- methods: numerical
 \end{keywords}

%%%%%%%%%%%%%%%%%%%%%%%%%%%%%%%%%%%%%%%%%%%%%%%%%%%

%%%%%%%%%%%%%%%%%%%%%%%%%%%%%%%%%%%%%%%%%%%%%%%%%%%
\section{Introduction} \label{sec:introduction}
%%%%%%%%%%%%%%%%%%%%%%%%%%%%%%%%%%%%%%%%%%%%%%%%%%%

Our Milky Way (MW) Galaxy is the best-studied galaxy in the
universe and provides some of the most stringent constraints of galaxy formation
models \citep[e.g.,][]{Guedes2011, Wetzel2016, Grand2017, Hopkins2018, Buck2020a}. Therefore, a thorough understanding of the components of the Galaxy, their dynamical evolution and formation channel is of great interest.

The Galaxy's main body, the rotationally-supported stellar disk harbors multiple components or populations when dissected in the space of chemistry, kinematics, spatial extent, and age \citep[e.g.,][]{Gilmore1983, Norris1985, Chiba2000, Nissen2010, Bovy2012, Haywood2013}; see \citet{Rix2013} and \citet{Bland-Hawthorn2016} for recent reviews.

Thanks to the great improvements in the quality and volume of astrometric datasets provided by the {\it Gaia} mission \citep{Gaia} in combination with chemical abundances and radial velocities from large spectroscopic surveys, (e.g. \textsc{apogee}; \textsc{galah}; \textsc{h3}; \textsc{LAMOST}; \citealt{Majewski2017}, \citealt{Martell2017,Buder2021}, \citealt{Conroy2019}, \citealt{Zhao2012}, respectively) we are now able to identify and link coherent chemodynamical structures in the MW to ancient merger events several billion years back in its history (e.g. \citealt{Helmi2020} for a review). By now a large number of chemodynamical clusters associated either with distinct in-situ components at the time of birth of the MW's stellar disk \citep[e.g.][]{Belokurov2022,Conroy2022,Rix2022} or associated with distinct accreted systems have been identified in the high-dimensional space of orbital parameters and stellar abundances \citep[e.g.,][]{Myeong2019, Naidu2020, Horta2021, Buder2022}.

Most notably, there exists an excess of stars on radial orbits in the local stellar halo around the Sun which is often referred to as {\it Gaia}-Sausage-Enceladus (GSE; \citealt{Belokurov2018, Helmi2018}; see also \citealt{Nissen2010, Koppelman2018, Haywood2018}). This stellar overdensity has been identified as the observable signature of a massive ($M_{\star} \simeq 10^{9}$ $\rm{M_{\odot}}$) early ($8-11$ Gyr ago; e.g. \citealt{Vincenzo2019, Belokurov2020, Naidu2021, Xiang2022}) accretion event. GSE is by far believed to be the most significant merger in MW's history and contributed approximately two-thirds of MW's stellar halo stars on highly-eccentric orbits \citep[e.g.][]{Mackereth2020}. GSE stars inhabit a sausage-like distribution in the radial-azimuthal velocity distribution \citep{Brook2003, Belokurov2018} and appear to be more metal-poor and less $\alpha$-enhanced than the redder halo counterpart \citep{Haywood2018, Helmi2018}. The redder sequence of the Galactic halo is thought to be the result of proto-galactic disc stars being dynamically ejected into the halo during the GSE merger and has been associated with an event termed {\it the Splash} by \cite{Belokurov2020, Bonaca2020} which has been initially found by other earlier studies \citep{Bonaca2017, Haywood2018, DiMatteo2019, Gallart2019}. 

The existence of the GSE merger and its strong impact on the structure of the proto-MW demands a thorough study of the processes that shape MW's chemo-dynamical evolution during such a merger in order to interpret the observational data at hand. In order to connect present day observables to Gyr old events we need tools like cosmological simulations that model galactic mass growth, mergers and subsequent star formation plus chemical enrichment self-consistently. Here, one prerequisite to follow the internal disk dynamics and suppress spurious heating as well as resolving the multi-phase, dense structure of the interstellar medium (ISM) is enough resolution ($N \gtrsim 10^6$; e.g. \citealt{Sellwood2013,Ludlow2019,Ludlow2021}). At the same time a proper model for chemical enrichment \citep[e.g.][]{Buck2021} is needed to accurately capture the star's birth kinematics (e.g. \citealt{Ratcliffe2022, Lu2022}) and the subsequent dynamical evolution of the disk (e.g. \citealt{House2011, Bird2013, Buck2020a}).

Thanks to recent advances in computing power and progress in numerical methods modern cosmological zoom simulations now meet the aforementioned requirements and are able to model individual MW-mass galaxies by sampling varying environments and formation scenarios (e.g. \citealt{Sawala2016, Grand2017, Buck2020a, Font2020, Applebaum2021, Agertz2021, Bird2021, Khoperskov2022-InSitu}). This enables us to reconstruct our Galaxy's formation history by linking the occurrence of chemodynamical patterns at $z=0$ to specific events in the Galaxy's evolution \citep[e.g.][]{Bignone2019, Fattahi2019, Grand2020, Elias2020, Dillamore2022, Khoperskov2022a, Rey2022}.

While GSE's strong impact on the MW's stellar halo is well established, more recently studies have further investigated its defining influence on the formation of the MW's thin and thick disk \citep[e.g.][]{Grand2020,Ciuca2022,Orkney2022,Rey2023}. These studies suggest that the GSE merger strongly shaped the formation of the MW's thick disk, agreeing with thick disk formation scenarios from early simulation results by \cite{Brook2004, Brook2006}. The gas-rich merger has a two-fold effect on MW's stellar disk: (i) it heats part of the existing proto-disc stars, dynamically ejecting them into the halo and creating the Splash. (ii) The merger provides fresh gas to the central galactic regions, triggering a starburst \citep[potentially after first causing a pause in star formation;][]{Vincenzo2019} that eventually forms the younger thick disc. After the merger, the thin disc then forms from the accretion of metal-poor gas in an inside-out, upside-down fashion \citep[e.g.,][]{Bird2013, Minchev2013, Grand2018, Buck2020a}. Using stellar populations around the solar radius from the APOGEE DR14 data \citet{Ciuca2021} qualitatively confirmed this picture. 

It has recently become possible to infer the birth radii, $R_{\rm{b}}$, of stars in the Milky Way in a model-free manner. \cite{Minchev2018} devised a simple new technique for obtaining $R_{\rm{b}}$ in any observational dataset with precise age and metallicity measurements. This method simultaneously recovered $R_{\rm{b}}$ and the Galactic disk metallicity evolution with radius and time, [Fe/H]$(R,\tau)$, by varying the metallicity slope as function of time so as to keep the resulting $R_{\rm{b}}$ distributions physically meaningful. More recently, \cite{Lu2022b} found that a linear relation exists between the metallicity scatter and the metallicity birth gradient evolution with cosmic time, thus making the $R_{\rm{b}}$ determination fully self-consistent. The results of the above work have opened up a new way of studying the chemo-dynamical evolution of the Milky Way, where instead of relying on forward modeling over cosmic time (cosmological simulation or semi-analytical modeling), one gets there directly from the data with few prior assumptions on the detailed chemical enrichment.

\begin{figure*}
    \script{merger_ratios.py}
    \begin{centering}
        \includegraphics[width=\linewidth]{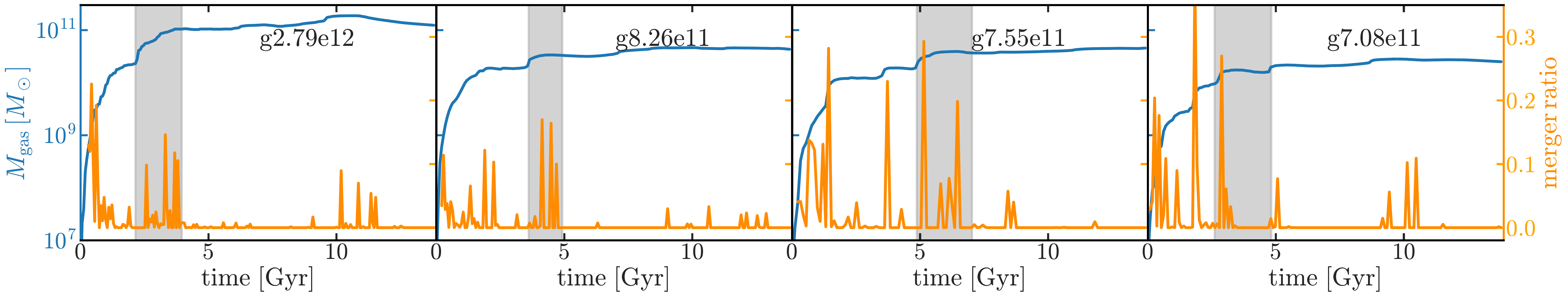}
        \vspace*{-1.75em}
        \caption{Gas mass (blue line, left axis) and gas merger ratio (orange line, right axis) as a function of cosmic time. The gray shaded areas mark the times when mergers heavily impact the gas disk size (see also Fig.~\ref{fig:half_mass}) chosen such that they bracket the time when the satellites cross the virial radius of the main galaxy and the time the satellite has fully coalescent with the main galaxy. Most of these mergers are at early times (2-4 Gyr), with the exceptions of g7.08e11 and g2.79e12, which also have a late time merger each.}
        \label{fig:merger_ratio}
    \end{centering}
\end{figure*}

Interestingly, when inferring the birth radii of MW disk stars, \citet{Lu2022b} found a steepening of the metallicity gradient at the time of the GSE merger using data from LAMOST DR7 combined with {\it Gaia} eDR3. When evaluating the chemical evolution across birth radii and lookback time using APOGEE DR17 red giant branch disk stars, \cite{Ratcliffe2023} reported a similar steepening in the metallicity gradient at $\sim9$ Gyr ago, with a similar trend also appearing in other [X/H] abundances. Here we use the NIHAO-UHD suite of cosmological hydrodynamical simulations of MW-mass galaxies \citep{Buck2020a} to investigate the influence of early merger events on the chemo-dynamics of the stellar disk. By using 4 different simulations from the NIHAO-UHD suite we investigate independently sampled environments that follow 4 different formation scenarios. We focus especially on the frequency of massive early mergers and their detailed impact on the formation of the stellar disk. This paper is structured as follows: In \S~2 we describe the simulations on which we base our analysis, followed by a presentation of the impact of early massive mergers on the metallicity gradient of the cold gas disk in \S~3. In \S~4 we turn to analyse where merger stars and gas end up in the age-metallicity relation  and the [$\alpha$/Fe] vs. [Fe/H] plane of the main galaxy. In \S~5 and \S~6 we summarize and conclude our findings.

%%%%%%%%%%%%%%%%%%%%%%%%%%%%%%%%%%%%%%%%%%%%%%%%%%%
\section{Simulations} \label{sec:simulation}
%%%%%%%%%%%%%%%%%%%%%%%%%%%%%%%%%%%%%%%%%%%%%%%%%%%

For this work we use four simulations from the NIHAO-UHD suite \citep{Buck2020a} part of the Numerical Investigation of a Hundred Astronomical Objects (NIHAO) simulation suite \citep{Wang2015}. These 4 galaxies are chosen to reflect the most MW like galaxies in terms of mass, size and diks properties. Parts of this simulation suite have previously been used to study the build-up of MW's peanut-shaped bulge \citep{Buck2018,Buck2019b}, investigate the stellar bar properties \citep{Hilmi2020}, infer the MW's dark halo spin \citep{Obreja2022}, study the dwarf galaxy inventory of MW mass galaxies \citep{Buck2019} or investigate the age-metallicity relation of MW disk stars \citep{Lu2022} including the chemical bimodality of disk stars \citep{Buck2020}, their abundances \citep{Lu2022a} and the origin of very metal-poor stars inside the stellar disk \citep{Sestito2021}.
Comparing the properties of these galaxies with observations of the MW and local disk galaxies from the SPARC data \citep{Lelli2016}, \citet{Buck2020a} showed that simulated galaxy properties agree well with observations.

\begin{figure*}
    \script{half_mass_radius_metal_gradient.py}
    \begin{centering}
     \includegraphics[width=\linewidth]{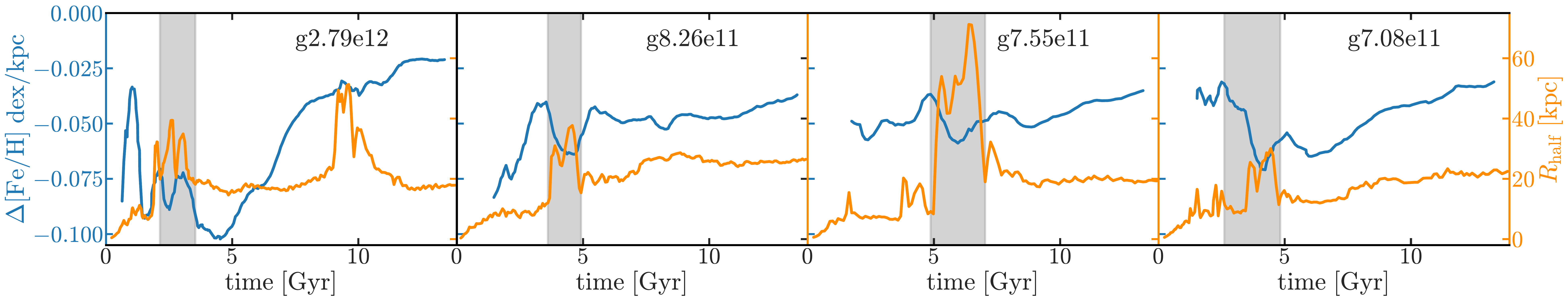}
        \vspace*{-1.75em}
        \caption{
            Half mass radius (orange line, right axis) and metallicity gradient (blue line, left axis) of cold ($T<$~15000~K) gas as a function of time. The sharp increases in the cold gas disc size around the times of gas rich mergers coincide with a sudden steepening of the metallicity gradient of $\sim0.02$ dex/kpc. The addition of relatively unenriched cold gas by the merging satellite leads to a sudden increase of the star-forming cold gas mass in the outskirts of the galaxies while the central parts are unaffected. This steepens the [Fe/H] gradient. Gray shaded areas are the same as in Fig.~\ref{fig:merger_ratio} and mark the times at which early massive and gas rich mergers impact strongly the cold gas disk size.
        }
        \label{fig:half_mass}
    \end{centering}
\end{figure*}

The simulations assume cosmological parameters from the \cite{Planck}, namely: \OmegaM= 0.3175, \OmegaL= 0.6825, \Omegab= 0.049, H${_0}$ = 67.1\kms\Mpc$^{-1}$, \sig8 = 0.8344. Initial conditions are created the same way as for the original NIHAO runs \citep[see][]{Wang2015} using a modified version of the \texttt{GRAFIC2} package \citep{Bertschinger2001,Penzo2014}. The mass resolution of these simulations ranges between $m_{\rm dark}\sim1.5 - 5.1\times10^5 \Msun$ for dark matter particles and $m_{\rm gas}\sim2.8 - 9.4\times10^4 \Msun$ for the gas particles. The corresponding force softenings are $\epsilon_{\rm dark}=414 - 620$ pc for the dark matter and $\epsilon_{\rm gas}=177 - 265$ pc for the gas and star particles. However, the adaptive smoothing length scheme implies that $h_{\rm  smooth}$ can be as small as $\sim30$ pc in the disk mid-plane. Stellar particles are born with an initial mass of $1/3\times m_{\rm{gas}}$ and are subject to massloss according to stellar evolution models as detailed in \citet{Stinson2013}. 
The simulation setup, star formation and feedback implementations are described in detail in the introductory paper \citep{Buck2020a} but for completeness we summarise them below. 

Simulations are performed with the modern smoothed particle hydrodynamics (SPH) solver {\texttt{GASOLINE2}} \citep{Wadsley2017} including substantial updates to the hydrodynamics as described in \citet{Keller2014}. {\texttt{GASOLINE2}} implements cooling via hydrogen, helium, and various metal-lines following \citet{Shen2010} using look-up tables calculated with \texttt{cloudy} \citep[version 07.02;][]{Ferland1998} and including photo-heating from the \citet{Haardt2005} UV background\footnote{For details on the impact of the UV background on galaxy formation see the recent study by \citet{Obreja2019}}. Star formation proceeds in cold (T $< 15,000$K), dense ($n_{\rm  th}  >  10.3$cm$^{-3}$) gas and is implemented as described in \citet{Stinson2006}. \citet{Buck2019a} showed that with this kind of star formation model only a high value of $n_{\rm  th}>10$cm$^{-3}$ \citep[see also][for an extended parameter study]{Dutton2019,Dutton2020} is able to reproduce the clustering of young star clusters as observed in the Legacy Extragalactic UV Survey (LEGUS) \citep{Calzetti2015,Grasha2017}.

Following \citet{Stinson2013} two modes of stellar feedback are implemented: (i) the energy input from young massive stars, e.g. stellar winds and photo ionization, prior to any supernovae explosions, thus termed \textit{early stellar feedback} (ESF). This mode consists of the total stellar luminosity ($2 \times 10^{50}$ erg of thermal energy per $M_{\odot}$) of the entire stellar population with an efficiency for coupling the energy input of $\epsilon_{\rm ESF}=13\%$ \citep{Wang2015}. (ii) supernova explosions implemented using the blastwave formalism as described in \citet{Stinson2006} and making use of a delayed cooling formalism for particles inside the blast region following \citet{McKee1977} in order to account for the adiabatic expansion of the supernova.
Finally, we adopted a metal diffusion algorithm between particles as described in \citet{Wadsley2008}.

The haloes in the zoom-in simulations were identified using the MPI+OpenMP hybrid halo finder \texttt{AHF2} \citep{Knollmann2009} and we use the accompanying merger tree tool to trace the particle IDs of all dark matter particles through time and identify all progenitor halos of a given galaxy/dark matter halo at redshift $z=0$. Subsequent analysis of the merger tree files is then performed with the {\texttt {ytree}} package \citep{ytree}. 

\begin{figure}
    \script{gas_profile_all.py}
    \begin{centering}
        \includegraphics[width=\linewidth]{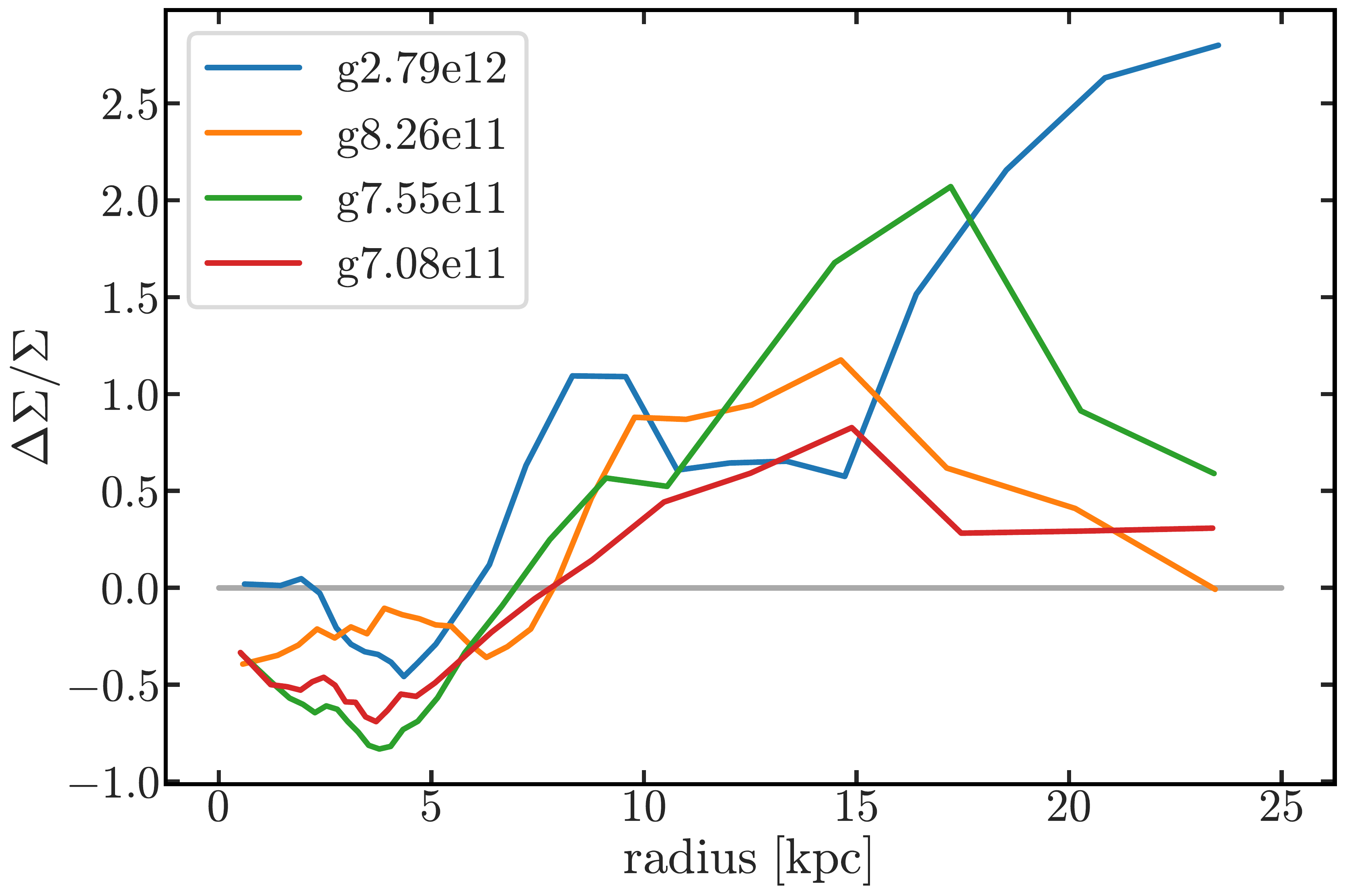}
        \vspace*{-1.75em}
        \caption{
            Relative change in cold gas surface density profile before and after the merger events as indicated in Fig.~\ref{fig:merger_ratio} by the gray shaded areas. Outside of a radius of $\sim5-10$ kpc the merging satellites contribute significant amounts of cold gas up to 200\%.
        }
        \label{fig:surf_den}
    \end{centering}
\end{figure}

%%%%%%%%%%%%%%%%%%%%%%%%%%%%%%%%%%%%%%%%%%%%%%%%%%%
\section{Impact on the cold gas metallicity gradient} \label{sec:results-gradient}
%%%%%%%%%%%%%%%%%%%%%%%%%%%%%%%%%%%%%%%%%%%%%%%%%%%

\subsection{The merger history} \label{sec:mergerhist}

\begin{figure*}
    \script{enrichment_evolution.py}
    \begin{centering}
        \includegraphics[width=\linewidth]{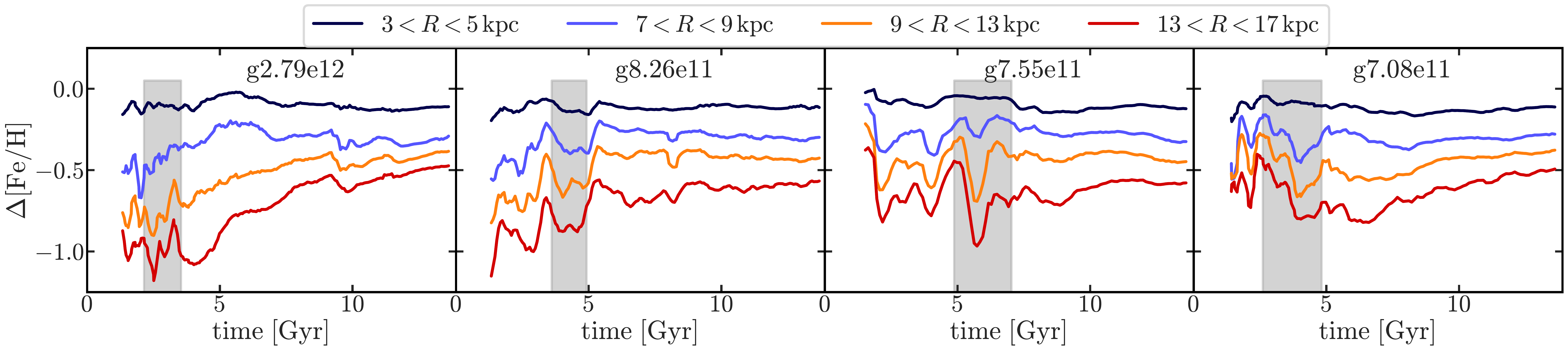}
        \vspace*{-1.75em}
        \caption{
            Cold gas metallicity difference $\Delta\mathrm{[Fe/H]}=\mathrm{[Fe/H]}(R)-\mathrm{[Fe/H]}(R<2\, \mathrm{kpc})$, between the gas in a given annulus at a radius $R$ (colored lines) and the metallicity in the central parts ($R<2$ kpc) of the galaxy, $\mathrm{[Fe/H]}(R<2\, \mathrm{kpc})$ as a function of time. A flat line indicates perfect self-similar evolution between two different radii while a more negative value indicates that a certain radius lacks behind the central galaxy in [Fe/H] enrichment and is thus indicative of a steepening of the metallicity gradient. Gray shaded areas mark times where the gas disk size sharply increases due to gas rich mergers (see Fig.~\ref{fig:half_mass}).
        }
        \label{fig:feh_evolution}
    \end{centering}
\end{figure*}

We start our analysis on the influence of Gaia-Sausage-Enceladus-like merger
events on the chemo-dynamics of MW-like galaxy disks by looking at the merger history of the four galaxies. Figure~\ref{fig:merger_ratio} shows the cumulative gas mass growth (blue line, left axis) accompanied by the gaseous merger ratio (orange line, right axis) defined as the ratio of gas mass of the merging satellite over the gas mass of the main galaxy measured at the time of the merger. We see that after an initial phase ($0-\sim2.5$ Gyr) of rapid gas mass growth accompanied by violent, relatively gas rich mergers there are several other mergers at later times that contribute more than 10\% in gas (as measured at the time of merging) leading to sudden jumps in the gas mass of the main galaxy (blue line). 
Especially for g2.79e12 (left most panel) and g7.08e11 (right most panel) we see some late time ($\sim10$ Gyr) gas rich merger while g8.26e11 and g7.55e11 do not show any gas rich merger after $5$ and $7$ Gyr, respectively. For the purpose of this paper, we highlight important gas rich merger events i.e. those that lead to a strong increase in the cold gas disc size (see also Fig.~\ref{fig:half_mass}) by gray shaded areas in Figs.~\ref{fig:merger_ratio}, \ref{fig:half_mass} and \ref{fig:feh_evolution}. Note that there might be more gas rich mergers especially at early times but our focus here is on the particular mergers that heavily impact the cold gas disk size.

\subsection{The impact of gas rich mergers on the metallicity gradient} \label{sec:merger}

The sudden increase of gas mass due to these gas rich mergers leads to a sudden increase in the half mass radius ($\gsim20$ kpc) of the cold gas disk of the main galaxy as the orange lines (right axis) in Figure~\ref{fig:half_mass} show. On the same figure we also plot the metallicity gradient of the cold gas disk of the main galaxy as a function of time (blue lines, left axis), measured by fitting a straight line to the radial metallicity profile in the radial range $2.5 - 17.5$ kpc. We have chosen this radial range such that it encompasses the star forming disk to highlight the effect of gas accretion on the central disk. Note, this radial range is sometimes smaller than the cold gas $R_{\rm half}$ during times of the merger which can be biased high by a non-spherical gas distribution like bridges etc (see e.g. Fig.~\ref{fig:visu}). In order to minimize short time scale fluctuations we plot the running average with a window size of 10 corresponding to a time window of $\sim600$~Myr. 
%One may raise question to whether a linear fit is sufficient during merger events, \cite{Lu2022Rblim} has investigated this and show that after the stellar disk has formed, there always exist a strong linear correlation between [Fe/H] and R.

We find that the time evolution of the metallicity gradient in the galaxies shows a rather complex behaviour. Galaxies g2.79e12 and g8.26e11 show a fast flattening of the initial metallicity gradient at early times ($\lesssim2$ Gyr and $\lesssim 2.5$ Gyr) from initially $-0.08$ dex/kpc to $-0.04$ dex/kpc while g7.55e11 and g7.08e11 start out already with a shallow gradient of $\sim-0.05$ and $\sim-0.04$.
Common to all four galaxies we find a steepening of the metallicity gradient by at least $0.02$ dex/kpc over a time frame of $\sim1$ Gyr at times where the gas disk size rapidly increases (marked with gray shaded areas) due to the increasing gas mass from the merging satellites. After the steepening happened, the gradient flattens again by $\sim0.01-0.02$ dex/kpc over roughy the same time frame except for g2.79e12 for which the gradient continues to steepen after the merger up to a maximum of $\sim-0.1$ dex/kpc. Only after another $\sim1.5$ Gyr the gradient starts to quickly flatten again until the present day. The late time evolution of the gradient of all four galaxies is comparatively weak but it consistently continues to flatten with at most a decrease of the gradient of $\sim0.02$ dex/kpc over a time of $\sim6-7$ Gyr.

\begin{figure*}
    \begin{centering}
    \hspace*{.5em} $\mathrm{all\, gas}$ \hspace*{12.5em} $\mathrm{[Fe/H]}>-1$ \hspace*{12.em} $\mathrm{[Fe/H]}<-1$ 
     \includegraphics[width=.99\linewidth]{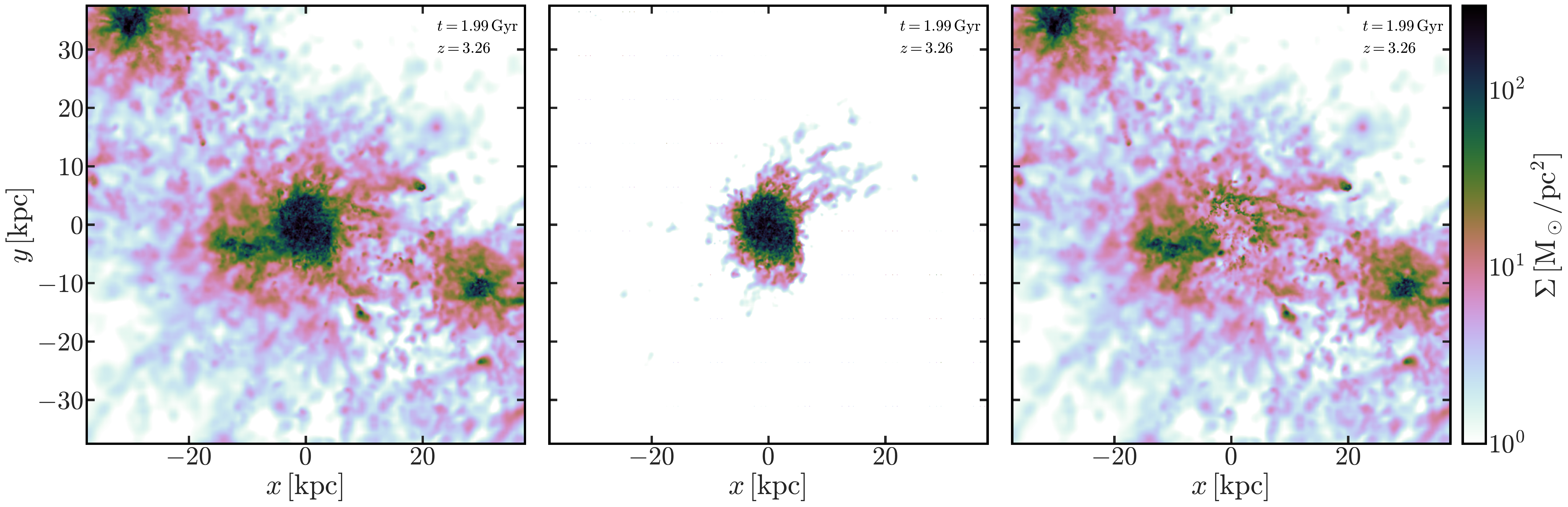}
     \includegraphics[width=.99\linewidth]{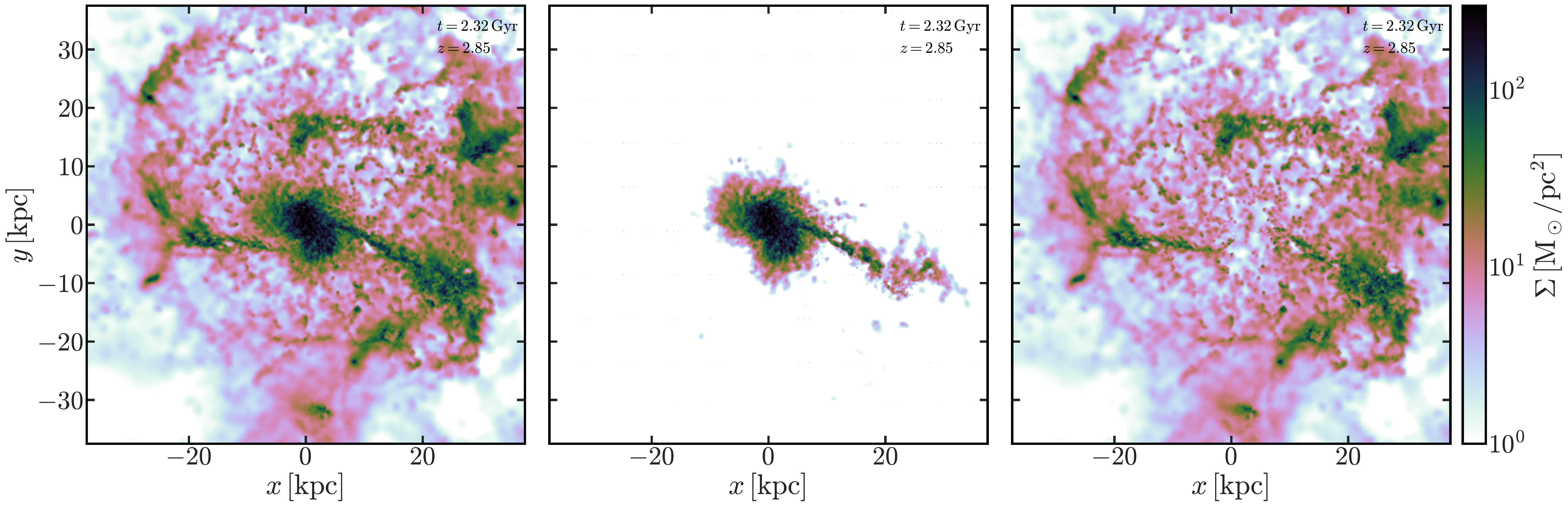}
     \includegraphics[width=.99\linewidth]{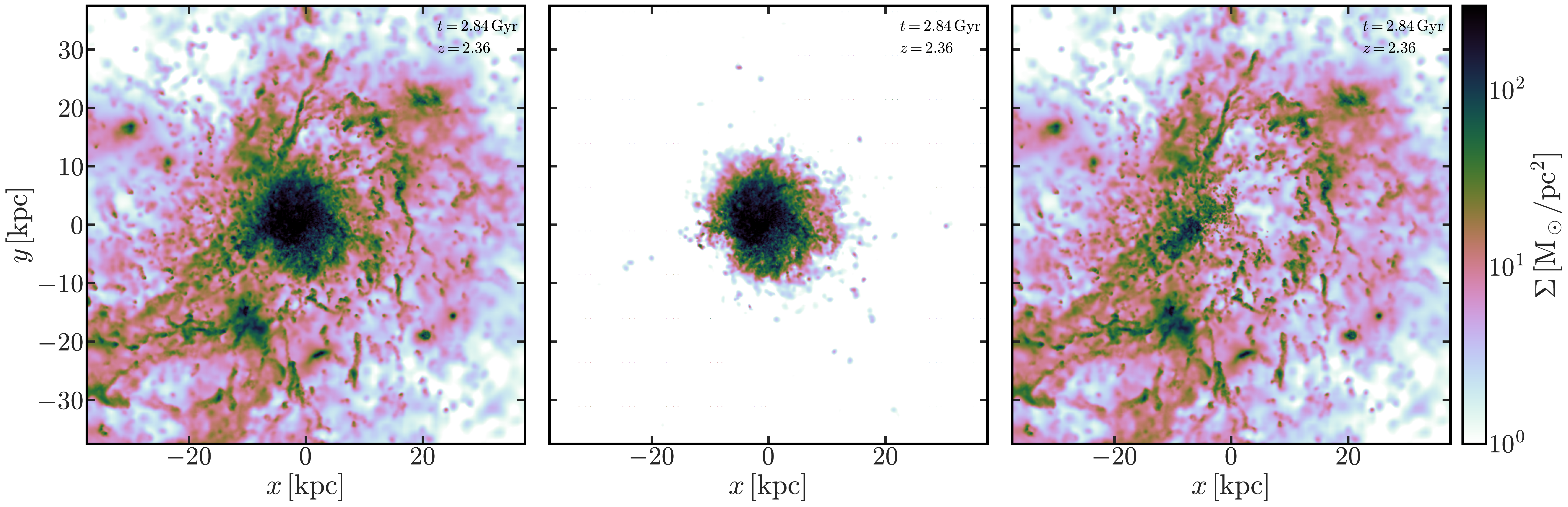}
     \includegraphics[width=.99\linewidth]{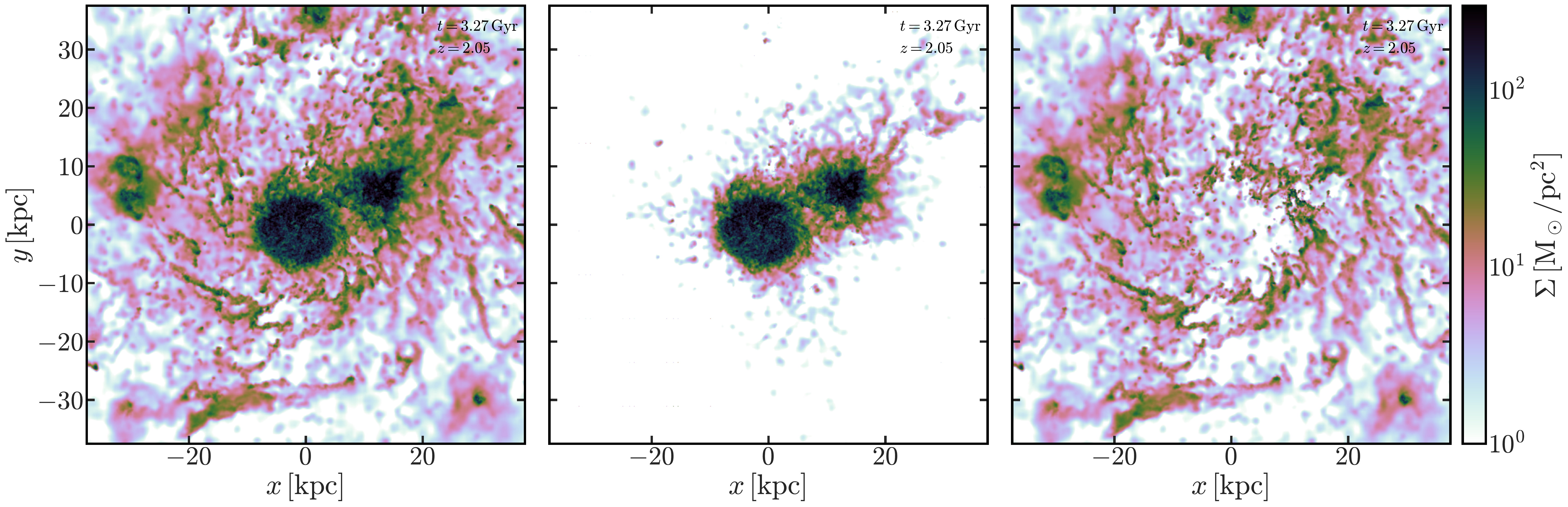}
     \vspace*{-1.em}
     \caption{
      Visualization of the cold gas surface density in galaxy g2.79e12 during the merger phase ($1.99\, \mathrm{Gyr} < t < 5.10\, \mathrm{Gyr}$; top to bottom, continued in Fig.~\ref{fig:visu2}) where the cold gas half mass radius sharply increases. From left to right we show the total cold gas surface density, the enriched ($\mathrm{[Fe/H]}>-1$) cold gas surface density and the unenriched cold gas ($\mathrm{[Fe/H]}<-1$). In the upper right corner of each panel we show the time and redshift.     
     }
     \label{fig:visu}
    \end{centering}
\end{figure*}

\begin{figure*}
    \begin{centering}
    \hspace*{.5em} $\mathrm{all\, gas}$ \hspace*{12.5em} $\mathrm{[Fe/H]}>-1$ \hspace*{12.em} $\mathrm{[Fe/H]}<-1$ 
     \includegraphics[width=.99\linewidth]{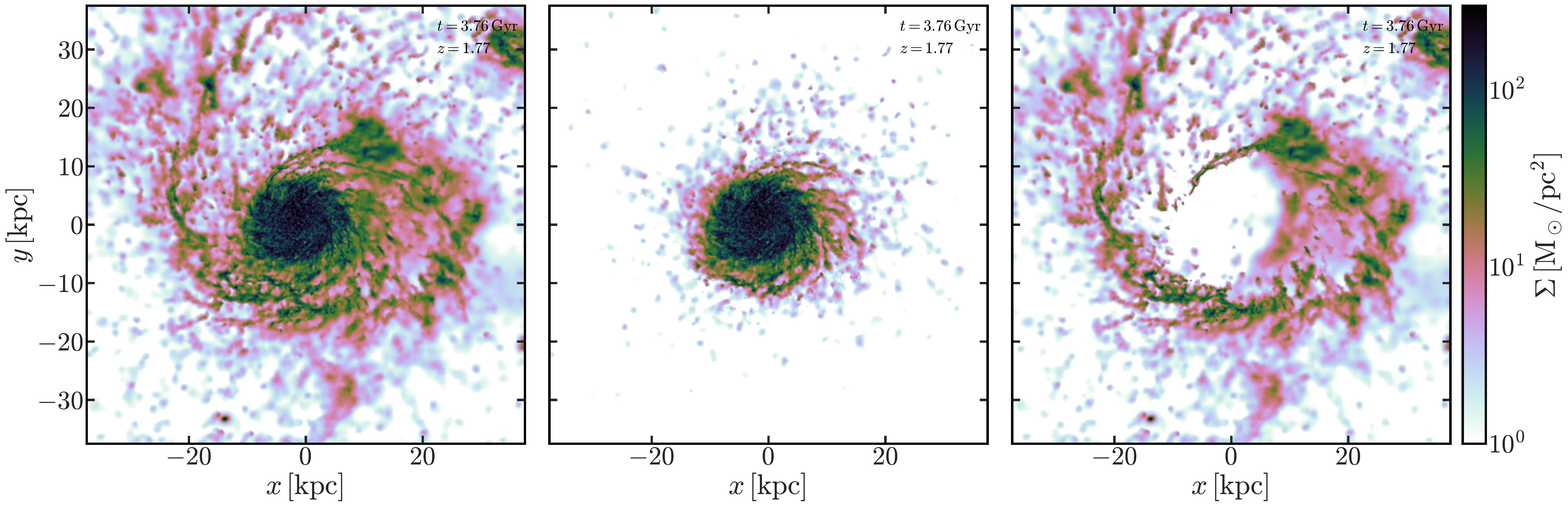}
     \includegraphics[width=.99\linewidth]{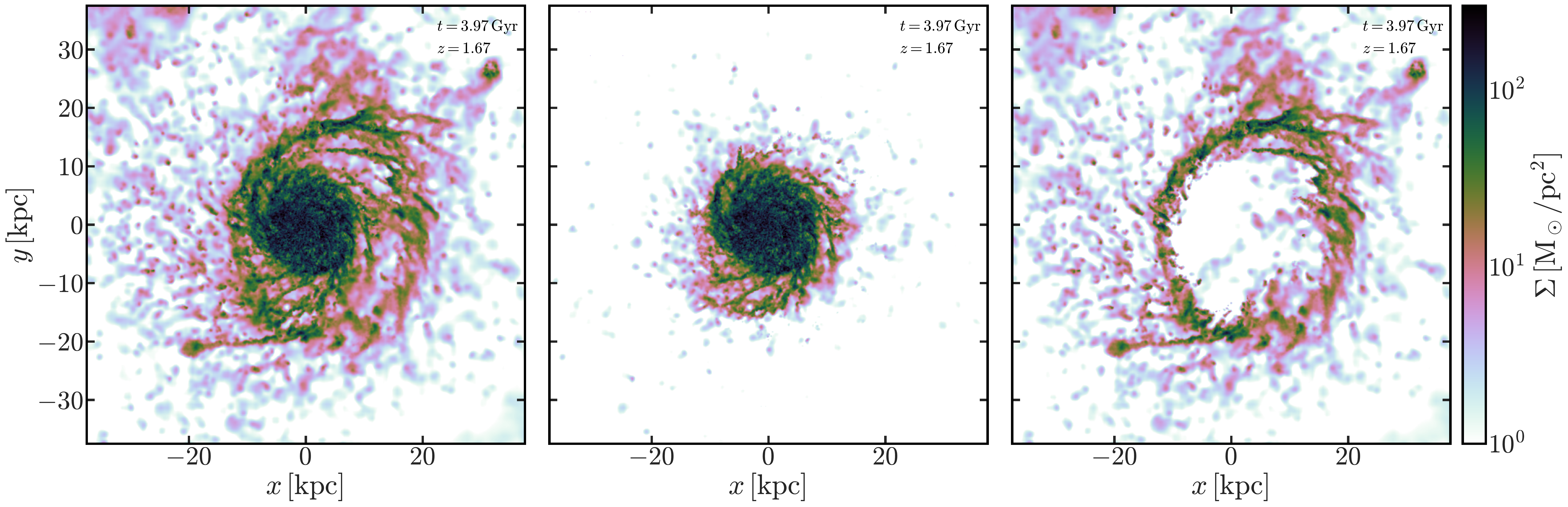}
     \includegraphics[width=.99\linewidth]{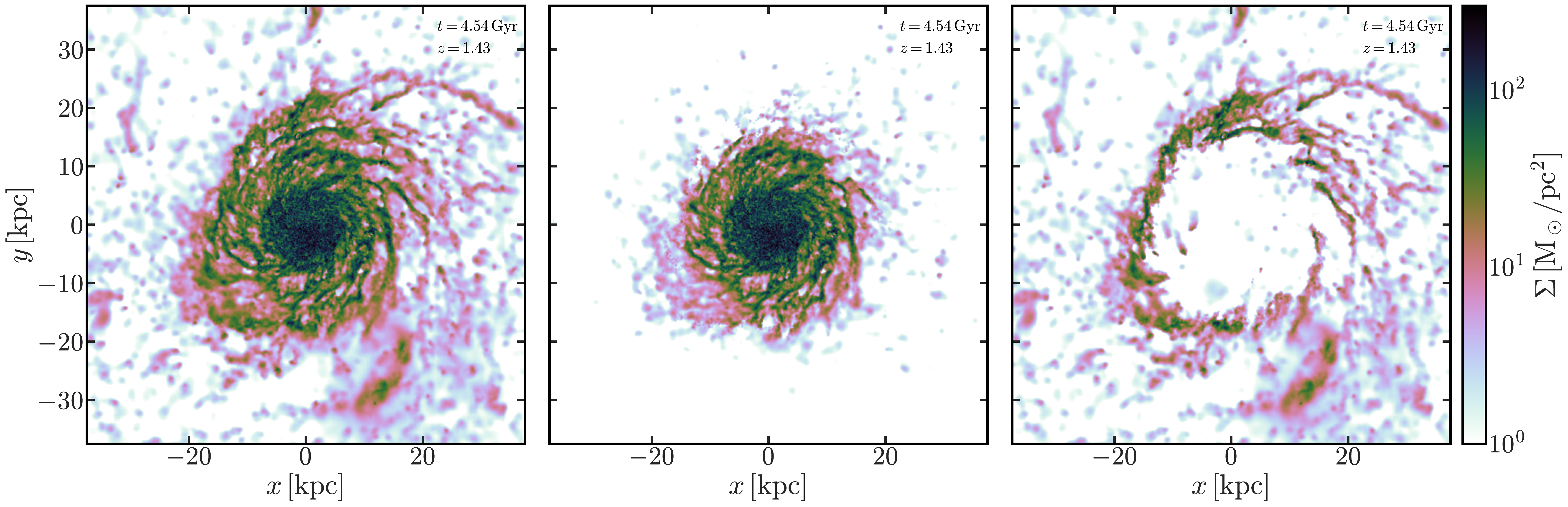}
     \includegraphics[width=.99\linewidth]{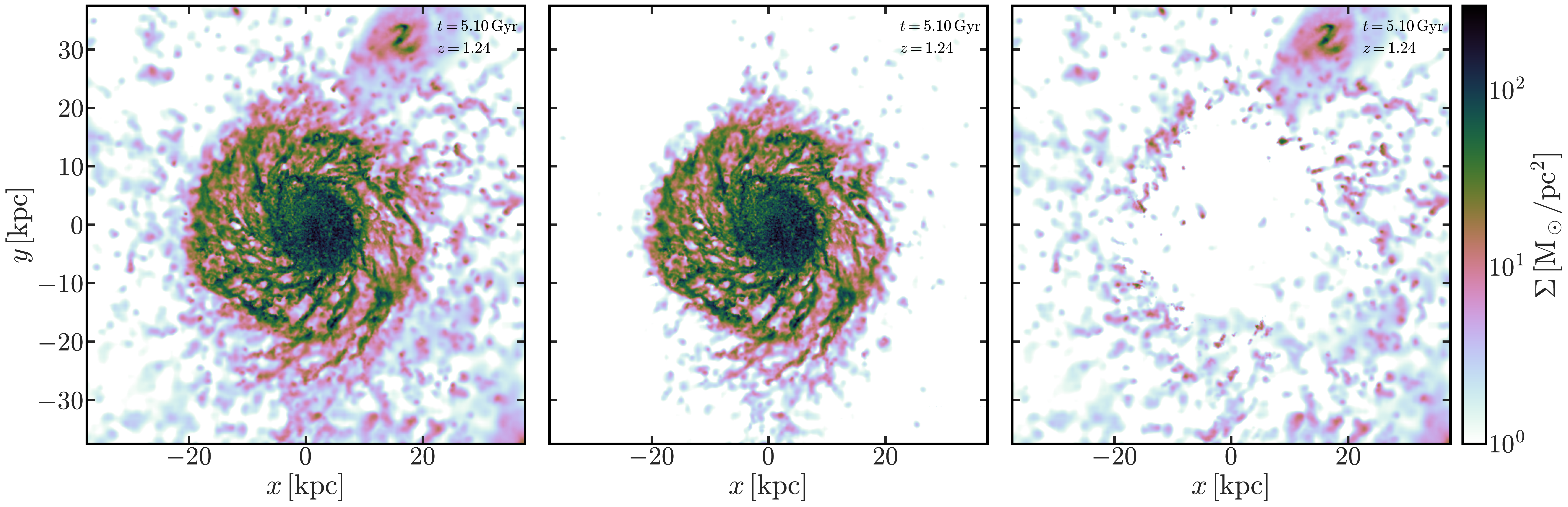}
     \vspace*{-1.em}
     \caption{
     Continuation of Fig.~\ref{fig:visu}.      
     }
     \label{fig:visu2}
    \end{centering}
\end{figure*}

\subsection{The physical mechanism behind a steepening of the metallicity gradient}
\label{sec:steepening}

A sudden steepening of the galactic metallicity gradient is unexpected and has (to our knowledge) been deemed not possible or at least not studied \citep[e.g.][for an extensive study]{Chiappini2001}. The physical reason for a steepening of the metallicity gradient might be (i) a dilution of the outer disk metallicity by the fresh, un-enriched gas brought in by the merging satellite, (ii) the continued or preferential enrichment of the galactic center, or (iii) a combination of both. A dilution scenario might be expected from the general lower metallicity of low mass satellites compared to MW-mass galaxies owing to the shape of the stellar mass-metallicity relation and has previously been suggested for the creation of the chemical bimodality of the MW's stellar disk \citep[e.g.][]{Chiappini1997, Spitoni2019, Buck2020}.

In order to investigate the cause of the strong steepening of the metallicity gradient in our simulations we first plot in Figure~\ref{fig:surf_den} where in the disk the accreted gas mostly contributes. Figure~\ref{fig:surf_den} plots the relative change of cold gas surface mass density radial profile before the gas rich merger and after the merger finished as indicated by the gray shaded areas in Figure~\ref{fig:half_mass}. This figure reveals an interesting finding. Consistently across all four galaxies the surface mass density of cold gas in the inner $5$ kpc is roughly constant or decreases slightly due to gas consumption by star formation and gas expulsion by feedback. Outside of $5$ kpc the surface mass density increases by $\sim75$ \% up to $\sim200$\% due to cold gas added by the merger event, either by direct accretion of cold gas from the merging satellite or by triggered gas accretion from the circum-galactic medium (CGM) due to tidal forces exerted by the interacting satellite. Note, the increase fraction correlates strongly with radius. Accretion of less enriched gas from the hotter CGM might be due to cold gas coagulation (whose seed can be the cold gas from the merging satellites or fountain gas \citep[e.g.][]{Armillotta2016,Sparre2020} from previous SNe/stellar winds) triggered by (gravitational) perturbations \citep[e.g.][]{Gronke2022}.

In fact, as was shown in \citet{Buck2020} in their Fig. 8 scenario (iii) is happening in these galaxies. For most of the time the ISM/cold gas metallicity evolution is self-similar and almost monotonically increasing at each radius but with markedly dilution events that correlate with merger events \citep[see also][]{Sparre2022}. In order to investigate this in more depth and study its effect on the metallicity gradient, we plot in Fig.~\ref{fig:feh_evolution} the difference in metallicity of the cold gas, $\Delta\mathrm{[Fe/H]}=\mathrm{[Fe/H]}(R)-\mathrm{[Fe/H]}(R<2\, \mathrm{kpc})$, between the gas in a given annulus at a radius $R$ and the metallicity in the central parts ($R<2$ kpc) of the galaxy, $\mathrm{[Fe/H]}(R<2\, \mathrm{kpc})$ as a function of time. Thus, flat lines in Fig.~\ref{fig:feh_evolution} would indicate perfect self-similar evolution and thus, no gradient change while a more negative value indicates that a certain radius lacks behind the central galaxy in [Fe/H] enrichment and thus a steepening of the metallicity gradient. On the other hand, a progressively more positive or less negative value shows that the radius at hand enriches more strongly in metallicity than the central parts and thus indicates a flattening of the (negative) metallicity gradient. In this way, we highlight deviations from the self-similar evolution which is indicating a steepening or flattening of the metallicity gradient as measured over different radial ranges, and we are able to disentangle whether it is the outskirts or the central parts that dominate the changes in the gradient. 

\begin{figure*}
    \script{mdf.py}
    \begin{centering}
        \includegraphics[width=\linewidth]{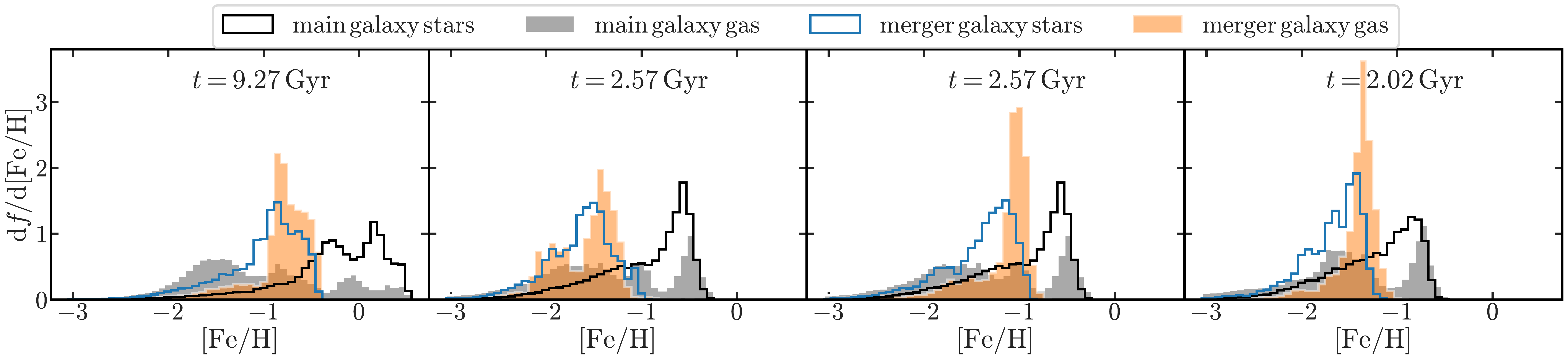}
        \vspace*{-1.75em}
        \caption{
            Metallicity distribution function of the galaxy g2.79e12 and the merging satellite shortly before coalescence for all four merger events, two happen at time $2.57$ Gyr. Filled histograms show the gaseous MDF while steps show the MDF for stars. In all cases, the gas metallicity of the satellite is $\sim0.5-0.75$ dex lower that the main galaxy's gas metallicity.
        }
        \label{fig:mdf}
    \end{centering}
\end{figure*}

Figure~\ref{fig:feh_evolution} complements our findings from Fig.~\ref{fig:half_mass} and shows that during phases of strong steepening of the metallicity gradient (gray shaded areas) it is especially the outskirts of the galaxies (outside of the solar circle, $R>8$ kpc) that lack behind the central parts in [Fe/H] enrichment. In general, we find that the inner annulus of $3<R<5$ kpc (black line) evolves almost self-similar to the central galaxy and relatively unaffected by the mergers. The outer galaxy annuli (blue, orange and red lines) show strong negative deviations/dips at the times of mergers (compare also to Fig.~\ref{fig:merger_ratio}), or similarly at the times of strong increase in the cold gas half-mass radius (see Fig.~\ref{fig:half_mass}), which indicates that the cold [Fe/H]-enriched gas at these radii is strongly influenced by the more metal-poor gas brought in by the merging dwarf galaxies. 

We have visualized this quantitative finding from Fig.~\ref{fig:feh_evolution} by plotting the cold gas surface density during the time of the gas rich merger for galaxy g2.79e12 in Fig.~\ref{fig:visu} and Fig.~\ref{fig:visu}. The panels in Fig.~\ref{fig:visu} and Fig.~\ref{fig:visu} are chosen such that they bracket the times before the merger (i.e. $t=1.99$ Gyr), during the merger event and final coalescence ($t=2.32, 2.84, 3.27$ Gyrs; Fig.~\ref{fig:visu}) as well as for some time after it to show the decrease in cold gas surface density by gas consumption and feedback ($t=3.76, 3.97, 4.54, 5.10$ Gyrs, Fig.~\ref{fig:visu2}). In these figures we show in the left column the total cold gas surface density and split it into the enriched ($\mathrm{[Fe/H]}>-1$) and an unenriched ($\mathrm{[Fe/H]}<-1$) contribution in the middle and right columns where our separating metallicity of $\mathrm{[Fe/H]}=-1$ is motivated by the shape of the MDF in Fig.~\ref{fig:mdf}. This figure clearly reveals how the merging satellite brings in a huge portion of unenriched cold gas that mainly settles into the outskirts of the pre-existing gas disk as we have seen from Fig.~\ref{fig:feh_evolution}. In total, the elevated, unenriched gas surface in the galaxy's outskirt survives for about 2.5-3 Gyr until its gas is consumed or heated up.

We note that after the merger, the elevated star formation in the galaxy outskirts causes these radii to quickly catch up in [Fe/H] enrichment \citep[see also Fig. 7 in][]{Buck2020}. Furthermore, Fig.~\ref{fig:half_mass} also shows that the gradual flattening of the metallicity gradient at late times (second half of the age of the Universe) is mostly caused by preferential/elevated enrichment of the galaxy outside the solar circle. At the same time \cite{Minchev2012a} showed that gas is migrating even stronger than the young stars will do and thus even a small amount of gas migrating outward will bring up $\Delta\rm{[Fe/H]}$ in the outer disk a lot.

%%%%%%%%%%%%%%%%%%%%%%%%%%%%%%%%%%%%%%%%%%%%%%%%%%%
\section{Impact on the age-metallicity relation and the [$\alpha$/Fe] vs. [Fe/H] plane} \label{sec:results-age}
%%%%%%%%%%%%%%%%%%%%%%%%%%%%%%%%%%%%%%%%%%%%%%%%%%%

We now turn to investigate the question of where the accreted stars and gas of the mergers appear in the age-metallicity relation and the [$\alpha$/Fe] vs. [Fe/H] plane. To this end we focus on the galaxy g2.79e12 for which the steepening of the metallicity gradient is the strongest but results are representative also for the other galaxies.

\subsection{The metallicity distribution function}
\label{sec:mdf}

 Figure~\ref{fig:mdf} shows the metallicity distribution function (MDF) for all the gas (orange histogram) and stars (blue histogram) inside the virial radius of four gas-rich satellites that merge with the main galaxy at $2.02$ Gyr, $2.57$ Gyr (two mergers happening at this time) and $9.27$ Gyr in comparison to the MDF of gas (gray histogram) and stars (black histogram) in the central galaxy at the corresponding times. A similar figure for the [$\alpha$/Fe] distribution function is shown in Fig.~\ref{fig:mdf_oxygen} of the appendix. Focusing first on the stellar components (step histograms) we see that the stars contributed by the mergers have on average $\sim0.75$ dex lower metallicity compared to the stars of the main galaxy although at the lowest metallicities ($\lesssim-2.5$) the tails of both distributions overlap. Turning to the gaseous component (filled histograms) we find that (i) the gas and stars of the dwarf galaxies have rather similar metallicity distribution functions where the gas seems to have slightly higher metallicity by about $\sim0.1$ dex. The gaseous MDF of the main galaxy on the other hand shows two peaks separated by about $\sim1$ dex. Those two peaks represent the star forming gas disk at higher metallicities between $-1\lesssim \rm{[Fe/H]}\lesssim0$ and the gas of the CGM at metallicities lower than $-1$. Comparing the gas MDF of the merger to the one of the main galaxy we recover the findings from the stellar component; the merger has an MDF that peaks $\sim0.75$ dex below the MDF of the main galaxy's gas disc and thus broadly overlaps with the CGM gas peak of the main galaxy. It is exactly this difference in metal enrichment in combination with the preferential addition of the accreted gas to the galaxy outskirts that causes the steepenig of the metallicity gradient in the main galaxy's star forming gas by about $0.025$ dex kpc$^{-1}$ as observed in Fig.~\ref{fig:half_mass} by reducing the metallicity in the galaxy outskirts by $\sim0.25$ dex as Fig.~\ref{fig:feh_evolution} shows. Thus, our expectation is that the GSE merger in the MW must have had a similarly lower metallicity at time of merging compared to the MW's gas disc in order to achieve a steepening of the MW's metallcity gradient of $\sim0.03$ dex as measured by \cite{Lu2022b} and \cite{Ratcliffe2023}.

\subsection{The age-metallicity relation}
\label{sec:age_fe}

\begin{figure*}
    \script{age_feh.py}
    \begin{centering}
        \includegraphics[width=\linewidth]{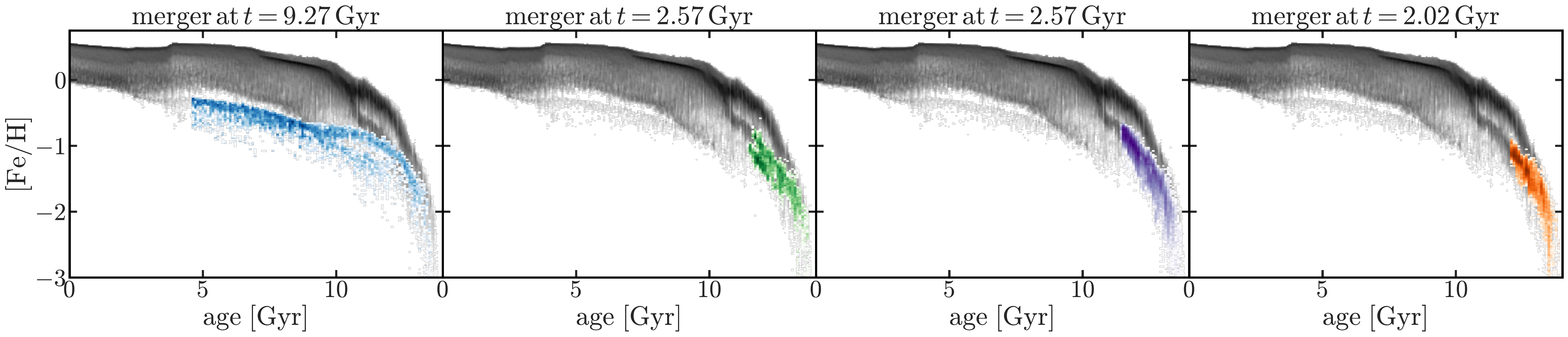}
        \vspace*{-1.75em}
        \caption{
            Age-metallicity relation for the stars in the galaxy g2.79e12 (black histogram in the background). Individual panels show with color where the stars contributed by the merging dwarf galaxies reside in age-metallicity space. The panel titles highlight the cosmic time at which the merger happens, with two different mergers happening at $t = 2.57$ Gyr.
        }
        \label{fig:age_feh}
    \end{centering}
\end{figure*}

Figure~\ref{fig:age_feh} shows the age-metallicity relation for stars of the same galaxy as discussed in the previous section (colored 2d-histograms) at a time shortly before merging with the main galaxy, in comparison to the age-metallicity relation of the main galaxy's stars at present day (black 2d-histogram).

Focusing first on the main galaxy, we see several prominent tracks/over-densities in the age-metallicity relation. The prominent upper envelop corresponds to the main galaxy. At large stellar ages we further see a second track at lower metallicity which is originating from the gas and stars accreted by the earlier mergers as is evident from the colored histograms showing the stars brought into the main galaxy by the merger. This second branch merges with the main galaxy's branch at around a stellar age of $\sim11$ Gyr, after all three early mergers have finished. That is shortly after the cold gas half mass radius of this galaxy rapidly increased (compare Fig.~\ref{fig:half_mass}), and the gas has finally mixed with the main galaxy's disk gas. Interestingly, as we have already seen in Fig.~\ref{fig:mdf}, the age-metallicity relation of the three early mergers does not differ too much and they largely overlap.

Finally, as already discussed in \cite{Buck2020}, the galaxy g2.79e12 suffered a massive late time merger, $\sim5$ Gyr ago The age-metallicity relation of this merger is shown in the leftmost panel of Fig.~\ref{fig:age_feh} and constitutes a third prominent and rather distinct branch in the age-metallicity relation that lies roughly $\sim0.5$ dex below the main galaxy's track. This track corresponds to the second rapid increase in the cold gas half mass radius as seen in Fig.~\ref{fig:half_mass} which stops the flattening of the metallicity gradient at late times. This track continues even to the youngest stars as the merger does two peri-center passages. Note the track of this merger itself splits into two tracks at stellar ages of $\sim10$ Gyr which indicates that the progenitor of this merger galaxy itself had an interaction in the early Universe.  Furthermore we see that at the time of the merger the entire age-metallicity relation of the main galaxy bends slightly down. Observational details of this bending have been discussed in \cite{Lu2022} closely resembling recent findings for the MW by \cite{Feuillet2019}. We will see in the next sub-section that this merger contributes a lot to the formation of the low-$\alpha$ sequence in this galaxy.

\subsection{The [$\alpha$/Fe] vs. [Fe/H] plane}
\label{sec:alpha_fe}

\begin{figure*}
    \script{feh_ofe.py}
    \begin{centering}
        \includegraphics[width=\linewidth]{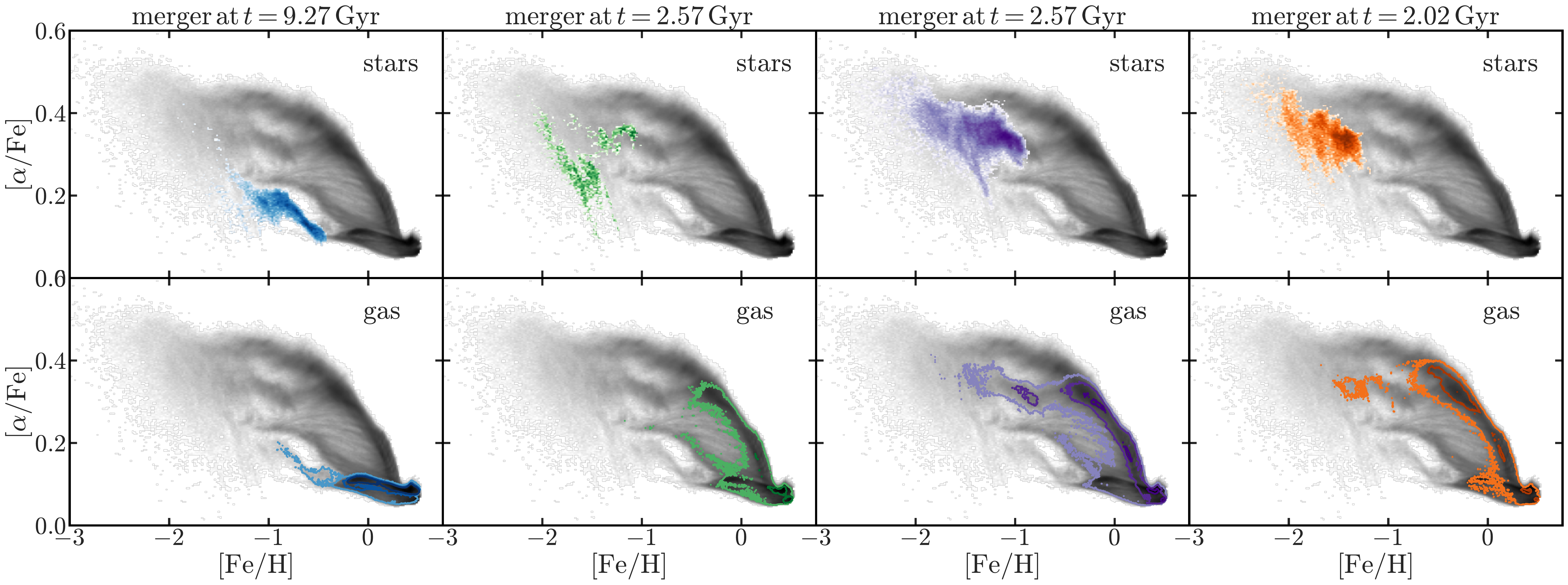}
        \vspace*{-1.75em}
        \caption{
            [$\alpha$/Fe] vs. [Fe/H] plane for the stars in the galaxy g2.79e12 (black histogram in the background). Individual panels show with color where the stars contributed by the merging dwarf galaxies reside in this plane (upper row), and where the gas from the merging dwarf galaxies contributes to star formation after the merger completed (lower panel). 
        }
        \label{fig:ofe_feh}
    \end{centering}
\end{figure*}

Figure~\ref{fig:ofe_feh} compares the [$\alpha$/Fe] vs. [Fe/H] plane of the main galaxy's stars (black 2d-histogram) to the stars (colored 2d-histogram in the upper row) and the stars born out of the gas (colored 2d-histogram in the bottom row) brought in by the merging dwarf galaxies discussed before.

Similarly to the age-metallicity relation we also find several tracks in the [$\alpha$/Fe] vs. [Fe/H] plane, especially at low metallicity. Again, the upper envelope track at the highest values of [$\alpha$/Fe] belongs to the main galaxy. There are several prominent tracks that run roughly parallel to this main track but at lower [$\alpha$/Fe] for fixed [Fe/H]. These are caused by the merging dwarf galaxies as a quick comparison to the colored histograms reveals. Additionally, at the lowest metallicities we find other faint stripes that run diagonal from high [$\alpha$/Fe] at lower metallicity to low [$\alpha$/Fe] at slightly higher metallicity. Those stripes originate from smaller dwarf galaxy mergers which are not the focus of this study. Again, we find that all three early mergers inhabit roughly the same loci in the [$\alpha$/Fe] vs. [Fe/H] plane as is expected from the MDFs discussed in the previous section. Interestingly the stars of those mergers only constitute the beginning of these secondary tracks.
Looking at the lower panels which show where the gas brought in by the merging dwarf galaxies contributes to the stars of the main galaxy we see that in fact the continued star formation from this merger gas continues the tracks until the gas finally completely mixes with the pre-existing gas in the main galaxy, to then contribute to the main track at the highest [$\alpha$/Fe] values for a given metallicity.

Finally, this figure clearly shows how the gas and stars of merging dwarf galaxies define their own tracks in the [$\alpha$/Fe] vs. [Fe/H] plane. In particular, these tracks are offset to lower [$\alpha$/Fe] and [Fe/H] values due to the fundamental mass-metallicity relation of galaxies \citep[e.g.][]{Kirby2013}. Especially, it clearly demonstrates how the late time merger at around $9.27$ Gyr introduces gradient fluctuations at each pericenter passage at times of $\sim10$ Gyr and $\sim12$ Gyr, and contributes significantly to the formation of the lower-$\alpha$ sequence in this galaxy. Such gradient fluctuations have been recently found by Ratcliffe at al. submitted and might be attributed to the recent interactions of the MW and the Saggitarius dwarf galaxy.

The observational implication of our results is that with low enough abundance uncertainties to not smooth out such features, we might be able to at least identify the most massive contributors to the MW (as has been demonstrated before e.g. \citealt{Vincenzo2019, Belokurov2020, Naidu2021, Xiang2022}) and gauge their effect on the MWs stellar body. Unfortunately, \cite{Ratcliffe2022} found that these distinguishing features vanished after the addition of present-day observational uncertainty. Further, our example of two merging dwarf galaxies at about the same time and inhabiting similar space in the [$\alpha$/Fe] vs. [Fe/H] plane (see Fig.~\ref{fig:age_feh}) shows the difficulty of disentangling one major merger from several mergers of lower mass. Especially current age uncertainties do not allow to separate multiple mergers that happen within $1-2$ Gyr nor do they allow to separate the two branches we see in the AMR in Fig.~\ref{fig:age_feh} for the proto-MW and the GSE-like mergers. This is particularly interesting as \cite{Donlon2022} recently suggested that the GSE feature might actually be created by multiple minor mergers. Thus, reducing the age uncertainties in future surveys might help to reveal this very interesting feature in MW data.

%%%%%%%%%%%%%%%%%%%%%%%%%%%%%%%%%%%%%%%%%%%%%%%%%%%
\section{Discussion}
\label{sec:discussion}
%%%%%%%%%%%%%%%%%%%%%%%%%%%%%%%%%%%%%%%%%%%%%%%%%%%

As stated previously, \cite{Lu2022a}, \cite{Ratcliffe2023} and Wang et al. submitted find a steepening in the stellar [Fe/H] gradient about $8-9$ Gyr ago. A steep metallicity gradient for such old stars seems surprising as the old, high-$\alpha$ stars in the MW do not show any significant metallicity gradient at present-day. It is now well established that [Fe/H] gradients of mono-age populations flatten from the effect of radial migration. This flattening is most prominent for the oldest stars, as shown, for example, by \citet[][their Fig.5]{Minchev2013}. Thus, while high-alpha stars have a flat [Fe/H] gradient today, there is a large scatter around it, related to the effect of radial migration \citep[see for example][Fig. 1]{Anders2017}. Thus, this steepening phase can only be revealed using the birth radius of stars as any kind of orbit reshuffling, like e.g. radial migration, will wash out the signal at present-day.

Here, using four hydrodynamical simulations, we find that early massive mergers can cause a steepening of the metallicity gradient, supporting their conclusions that the GSE was responsible. GSE is indeed expected to have been a major merger of about 4:1, or 20\% of the MW mass \citep{Helmi2018} and most likely gas-rich, as expected for mergers at high redshift. An attempt to estimate its gas fraction at the time of merging was done by \citep{Vincenzo2019}, finding a value of ~0.67.

One difference between our work here and the work of Ratcliffe et al. and \cite{Lu2022a} is that we find the gradient of the gas steepens for two out four galaxies for only $\sim$1 Gyr before beginning to flatten out again, while both \cite{Lu2022a} and \cite{Ratcliffe2023} find that the stellar metallicity gradient steepens over a time frame of 2+ Gyr. Such a long phase of steepening is only observed in the galaxy g2.79e12 and g7.08e11, for which then also the phase of flattening is longer compared to the other two galaxies in our sample and last from the time of merger until the present day. Understanding the roll of the relative merger mass size, velocity, and metallicity, as well as observational age and [Fe/H] uncertainties, on the impact of this time frame is beyond the scope of this work. Nevertheless, the conclusions drawn here highlight the effect of mergers on galactic disk metallicity gradients, and clarify that the [Fe/H] gradient in the MW disk does not monotonically flatten with time as assumed in some previous works (e.g. \citealt{Kubryk2015, Minchev2018}).

Figure~\ref{fig:age_feh} suggests that the merger history of the MW should leave a unique signature in its AMR. \cite{Ciuca2022} identify a ``dip" and a diagonal feature increasing in [Fe/H] with cosmic time which they call the ``Great Galactic Starburst" in the MW's denoised AMR. Our results support their conclusions that these features were caused by an early massive merger (GSE). They discuss that these features are different regions of the thick disk (high-$\alpha$ sequence). We now understand that this ``starburst" and dip episode are due to the merging of the second branch with the main galaxy's branch after the early merger finished.

In addition to the primary steepening phase just after an early massive merger, we find additional steepening phases in the [Fe/H] gradient in each simulation (Fig.~\ref{fig:half_mass}).  Since a large percentage of the gas is stripped during the first crossing of a merger (\citealt{TepperGarcia2018}), we would expect subsequent passages to have a weaker effect on the gradient. Looking at g2.79e12, which has a late time merger at $\sim$9.3 Gyr, with pericenter passages at $\sim$10 and 12 Gyr, we see that the first pericenter passage has a stronger effect on the metallicity gradient than the second passage (Fig.~\ref{fig:half_mass}). This mimics the findings in observational data by \cite{Ratcliffe2023}, who discovered three fluctuations in the metallicity gradient, and attributed the later two fluctuations to Sagittarius’ pericenter passages. If Sagittarius was the culprit, we would expect a second track below the main galaxy's in the AMR if there is a late time merger (Fig.~\ref{fig:age_feh}). This signature may be visible in some age catalogs \citep{Xiang2022,Anders2023}, however current age and metallicity uncertainties mask the distinctness of the two tracks. Still, our results suggest that Sgr may be responsible for these features both in the AMR and the metallicity gradient. Future work with precise and accurate stellar ages and metallicity could use these multiple steepenings in the metallicity gradient to constrain passage times and potentially the process of gas stripping of merging satellites.

Finally, \cite{Annem2022} find that the effect of gas infall from Sagittarius is primarily felt in the outer disk, where the gas is diluted in [Fe/H]. This agrees with our Fig.~\ref{fig:feh_evolution}, where we directly show that the dilution in [Fe/H] is felt more in the outer disk, and rarely felt near the galactic center.

%%%%%%%%%%%%%%%%%%%%%%%%%%%%%%%%%%%%%%%%%%%%%%%%%%%
\section{Conclusion}
\label{sec:conclusion}
%%%%%%%%%%%%%%%%%%%%%%%%%%%%%%%%%%%%%%%%%%%%%%%%%%%

Recent observations of the MW by \citet{Lu2022b} using data from LAMOST DR7 combined with {\it Gaia} eDR3 and \cite{Ratcliffe2023} using APOGEE DR17 surprisingly found a steepening of the metallicity gradient at the time of the GSE merger \citep{Belokurov2018, Helmi2018}. In this work we investigated the influence of early massive merger events on the evolution of the metallicity gradient of the cold star forming gas disk of simulated galaxies. We made use of the NIHAO-UHD suite \citep{Buck2020} of cosmological hydrodynamical simulations of MW-mass galaxies to study the frequency of massive early mergers (see Fig.~\ref{fig:merger_ratio}) and their detailed impact on the gaseous disks in the simulated galaxies. Our results are summarized as follows:
\begin{itemize}
\item All four galaxies used in this work exhibit a strong steepening of the metallicity gradient at early times, much like the observations of \citet{Lu2022b} and \cite{Ratcliffe2023} revealed for the MW (blue lines in Fig.~\ref{fig:half_mass}).  
\item The steepening of the cold gas metallicity gradient is caused by massive, gas rich early merger events that produce a sudden increase in the cold gas disk size (orange lines in Fig.~\ref{fig:half_mass}).
\item The merging dwarf galaxies mostly affect the galaxy outskirts and lead to an increase in cold gas surface density of up to 200\% (see Fig.~\ref{fig:surf_den}).
\item In more detail, the addition of relatively un-enriched, fresh gas by the mergers break the monotonous enrichment of the ISM and cause a dilution of the cold gas in the outskirts of the galaxies (see Fig.~\ref{fig:feh_evolution}) which in turn leads to a steepening of the metallicity gradient.
\item Investigating the metallicity distribution function (MDF) for gas and stars of the merging dwarf galaxies (Fig.~\ref{fig:mdf}) in comparison to the main galaxy at the time of merger we find that the mergers typically have MDFs that peak $\sim0.75$ dex below the main galaxy's MDF. This difference together with the preferential addition of cold gas to the main galaxy outskirts (see Fig.~\ref{fig:surf_den}) causes the strong steepening of the metallicity gradients in our simulations.
\item The stars accreted with the merging dwarf galaxies and the ones later formed out of the accreted gas inhabit distinct locations/tracks from the main galaxy's stars in the [$\alpha$/Fe] vs. [Fe/H] plane (see Fig.~\ref{fig:ofe_feh}), offset to lower [$\alpha$/Fe] and [Fe/H] values due to the fundamental mass-metallicity relation of galaxies. 
\item Such early mergers can contribute significantly to the formation of a second, low-$\alpha$ sequence in the [$\alpha$/Fe] vs. [Fe/H] plane as observed in the MW.  
\end{itemize}

%%%%%%%%%%%%%%%%%%%%%%%%%%%%%%%%%%%%%%%%%%%%%%%%%%%
\section*{Data Availability}
This manuscript has been typeset using showyourwork! by \citet{Luger2021} and thus is fully reproduceable. All data and code to recreate the analysis presented here can be found at \url{https://github.com/TobiBu/GSE_merger}. The simulations underlying this work are publicly available at \url{https://tobias-buck.de/#sim_data}.
%%%%%%%%%%%%%%%%%%%%%%%%%%%%%%%%%%%%%%%%%%%%%%%%%%%

%%%%%%%%%%%%%%%%%%%%%%%%%%%%%%%%%%%%%%%%%%%%%%%%%%%
\section*{acknowledgments}
%\section*{Acknowledgments}
%%%%%%%%%%%%%%%%%%%%%%%%%%%%%%%%%%%%%%%%%%%%%%%%%%%
TB's contribution to this project was made possible by funding from the Carl Zeiss Foundation. AO is funded by the Deutsche Forschungsgemeinschaft (DFG, German Research Foundation) –- 443044596. BR and IM acknowledge support by the Deutsche Forschungsgemeinschaft under the grant MI 2009/2-1. We gratefully acknowledge the Gauss Centre for Supercomputing e.V. (www.gauss-centre.eu) for funding this project by providing computing time on the GCS Supercomputer SuperMUC at Leibniz Supercomputing Centre (www.lrz.de). This research was carried out on the High Performance Computing resources at New York University Abu Dhabi.
This research made use of the {\sc{pynbody}} \citet{pynbody} package to analyze the simulations and used the {\sc{ytree}} package \citep{ytree} to analyze the AHF merger trees. We made use of the {\sc{python}} package {\sc{matplotlib}} \citep{matplotlib} to display all figures in this work. Data analysis for this work made intensive use of the {\sc{python}} library {\sc{SciPy}} \citep{scipy}, in particular {\sc{NumPy, IPython and Jupyter notebooks}} \citep{numpy,ipython,jupyter}. The article has been typeset using showyourwork! by \citet{Luger2021}. We thank Dan Foreman-Mackey and Rodrigo Luger for valuable help with showyourwork!.

\bibliography{bib.bib}

\appendix

%\section{Rotation velocity of stars}

%\begin{figure}
%    \script{v_phi_stars.py}
%    \begin{centering}
%        \includegraphics[width=\linewidth]{figures/2.79e12_v_phi_stars}
%        \caption{
%            Same as Fig.~\ref{fig:v_phi} but for the stellar rotation velocity in the disk plane, $v_\phi$, before and after the merger. The peak of the $v_\phi$ distribution shifts from 27 km/s before the merger to 66 km/s after the merger, more than doubling the rotation speed of the stellar disk.
%        }
%        \label{fig:v_phi_stars}
%    \end{centering}
%\end{figure}

\begin{figure*}
    \script{mdf_oxygen.py}
    \begin{centering}
        \includegraphics[width=\linewidth]{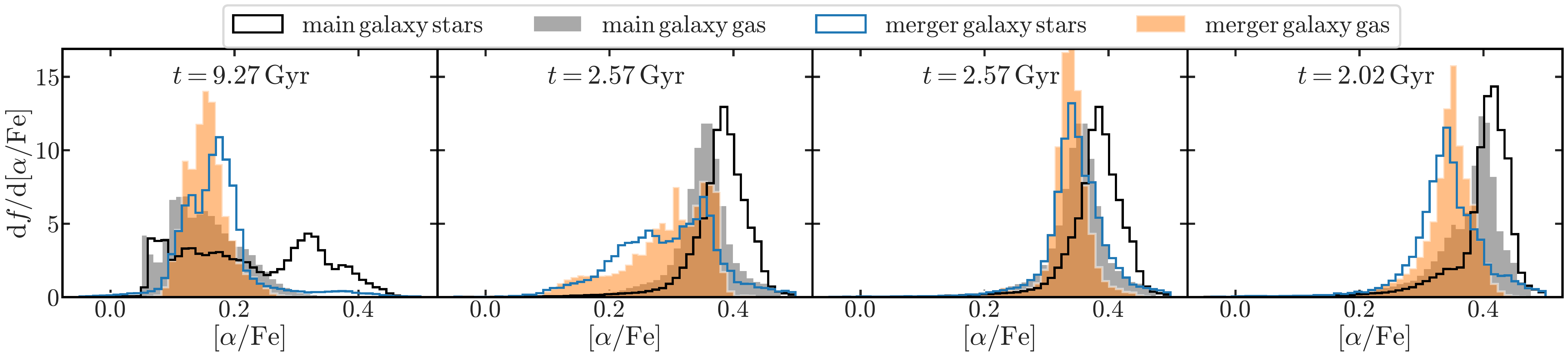}
        \vspace*{-1.75em}
        \caption{
            Same as Fig.~\ref{fig:mdf} but for the oxygen abundance. Filled histograms show the gaseous oxygen abundance distribution while steps show the one for stars. The gas oxygen abundance of the satellite is  $\sim0.1$ dex lower that the main galaxy's gas oxygen abundance.
        }
        \label{fig:mdf_oxygen}
    \end{centering}
\end{figure*}

\section{Oxygen abundance distribution}

Figure~\ref{fig:mdf_oxygen} compares the [$\alpha$/Fe] distribution function (MDF) for gas (orange histogram) and stars (blue histogram) of four gas rich satellites that merge with the main galaxy at $2.02$ Gyr, $2.57$ Gyr and $9.27$ Gyr in comparison to the MDF of gas (gray histogram) and stars (black histogram) in the central galaxy at the corresponding times.
This figure shows that the stellar [$\alpha$/Fe] (open histograms) of the merger stars is lower compared to the main galaxy's stars with largely varying offsets between $0.05$ and $0.2$ dex. Looking instead at the gas [$\alpha$/Fe] (filled histograms) we find that the difference between merger gas and main galaxy gas is less prominent. We find the largest difference for the merger at $2.02$ Gyr ($\sim0.05$ dex) and largely an overlap for the late time merger at $9.27$ Gyr. The two mergers at $2.57$ Gyr only show a mild offset of the merger gas to lower [$\alpha$/Fe] of $\lesssim0.02$ compared to the main galaxy although the second panel from left shows a tail towards lower [$\alpha$/Fe] by about $\sim0.1$ dex.

\label{lastpage}
\end{document}